# A self-managing fault management mechanism for wireless sensor networks

Muhammad Asim[1] and Hala Mokhtar[2] and Madjid Merabti[3]

[1] School of Computing and Mathematical Sciences, Liverpool John Moores University
`M.Asim@2006.ljmu.ac.uk`

[2] School of Computing and Mathematical Sciences, Liverpool John Moores University
`H.M.Mokhtar@ljmu.ac.uk`

[3] School of Computing and Mathematical Sciences, Liverpool John Moores University
`M.Merabti@ljmu.ac.uk`

## ABSTRACT

*A sensor network can be described as a collection of sensor nodes which co-ordinate with each other to perform some specific function. These sensor nodes are mainly in large numbers and are densely deployed either inside the phenomenon or very close to it. They can be used for various application areas (e.g. health, military, home). Failures are inevitable in wireless sensor networks due to inhospitable environment and unattended deployment. Therefore, it is necessary that network failures are detected in advance and appropriate measures are taken to sustain network operation. We previously proposed a cellular approach for fault detection and recovery. In this paper we extend the cellular approach and propose a new fault management mechanism to deal with fault detection and recovery. We propose a hierarchical structure to properly distribute fault management tasks among sensor nodes by introducing more 'self-managing' functions. The proposed failure detection and recovery algorithm has been compared with some existing related work and proven to be more energy efficient.*

## 1. INTRODUCTION

Fault management has been widely considered as a key part of today's network management. Recent rapid growth of interests in Wireless Sensor Networks (WSNs) has further strengthened the importance of fault management, or in particular, played a crucial role. Faults in WSNs are not exception and tend to occur more frequently. In addition to typical network faults, wireless sensor networks have to deal with faults arising out of unreliable hardware, limited energy, connectivity interruption, environmental variation and so on. Thus, in order to guarantee the network quality of service and performance, it is essential for WSNs to be able to detect failures and to perform something akin to heal and recover the network from events that might cause faults or misbehaviour. A set of functions and applications designed specifically for this purpose is called a fault management platform [1-3].

One way of dealing with faults is to design a system that is fault-tolerant to begin with. Fault tolerance is the ability to maintain sensor networks functionalities without any interruption due to sensor nodes failure. However, this requires network designer to be fully aware, at design time, of the different types of faults and the extent to which they may occur once the network is deployed. The power supply is the most critical restriction as it is usually difficult to be rechargeable. For this reason faults occurs frequently and will not be isolated events. Attacks by adversaries could happen because these networks will be often embedded in critical applications. Worse, attacks could be facilitated because these networks will be deployed in open spaces or enemy territories, where adversaries cannot only manipulate the environment but gain physical access to the node. Also, communication in sensor networks takes place by radio frequencies means that adversaries can easily inject themselves in the network and disrupt infrastructure functions. Moreover, sensor nodes are commonly used to monitor external environment, due to which sensor nodes are susceptible to natural phenomenons like rain, fire and fall of trees [4].

Sensor network faults cannot be approached similarly as in traditional wired or wireless networks due to the following reasons [2]:

1. Traditional wired network protocol are not concerned with the energy consumptions as they are constantly powered and wireless ad hoc networks are also rechargeable regularly.

2. Traditional network protocols aim to achieve point-to-point reliability, where as wireless sensor networks are more concerned with reliable event detection.

3. Faults occur more frequently in wireless sensor networks than in traditional networks, where client machine, servers and routers are assumed to operate normally.

In this paper, we extend our existing cellular architecture for fault detection and recovery [5] and describe a new fault management mechanism to detect failing nodes and recover the connectivity in WSNs. We propose a new fault knowledge model to support sensor nodes responding to network faults. Also, this paper attempts to examine the efficiency of our existing cellular architecture for fault detection and recovery. In our proposed cellular architecture, the whole network into a virtual grid of cells. A cell manager is chosen in each cell to perform management tasks. These cells combine to form various groups and each group chooses one of their cell managers to be a group manager. We propose using a hierarchical management structure to ensure that self-management ability is respectively distributed. The hierarchical management framework and node management role is also expected to be self-adjustable dynamically to the changes occurred in the network. For examples, replacing the failed cell manager; shifting over some workload from the sensor nodes whose residual resource status is in a critical level. The faulty sensor nodes are detected and recovered in their respective cells without affecting overall structure of the network. We also presented some simulation results to prove the efficiency of our cellular architecture.

## 2. RELATED WORK

Existing fault management approaches for WSNs vary in forms of architectures, protocols, detection algorithm or detection decision fusion algorithm etc [3]. A survey on fault tolerance in wireless sensor networks can be found in [2]. This section starts by reviewing the fault detection approaches, then we present fault diagnosis and failure recovery mechanisms.

### 2.1 Fault detection

Since sensor network conditions undergo constant changes, network monitoring alone may not be sufficient to identify network faults. Therefore, fault detection techniques need to be in place to detect potential faults [2]. Generally, fault detection in WSNs has two types: explicit detection and implicit detection [3]. The first one is performed directly by the sensing devices and their sensing applications. The implicit detection refers that anomalistic phenomena might disable a sensor node from communication or behave properly, and has to be identified by the network itself. Implicit detection is normally achieved in two ways: active and passive model. The active detection model is carried out by the central controller of sensor network. Sensor nodes continuously send keep-alive messages to the central controller to confirm their existence. If the central controller does not receive the update message from a sensor node after a pre-specified period of time, it may believe that the sensor is dead. Passive detection model (event-driven model) triggers the alarm only when failure has been detected. However this model will not work properly if a sensor is disabled from communication due to intrusion, tampering or being out of range. Fault detection mainly depends on the type of application and the type of failures. Some exiting fault detection schemes are discussed below. We classify the existing failure detection approaches into two primary types: centralized and distributed approach.

*A. Centralized approaches*

In centralized fault management systems, usually a geographical or logical centralized sensor node identifies failed or misbehaving nodes in the whole network. This centralized node can be a base station, a central controller or a manager. This central node usually has unlimited resources and performs wide range of fault management tasks [3]. Some common centralized fault management approaches are as follows:

Sympathy [6] is a debugging system and is used to identify and localize the cause of the failures in sensor network application. Sympathy algorithm does not provide automatic bug detection. It depends on historical data and metrics analysis in order to isolate the cause of the failure. Sympathy may require nodes to exchange neighbourhood list, which is expensive in terms of energy. Also, Sympathy flooding approach means imprecise knowledge of global network states and may cause incorrect analysis.

Jessica Staddon et al [7] enabled the base station to construct an overview of network by integrating each piece of network topology information (i.e. node neighbour list) embedded in node usual routing message. This approach uses a simple divide-and-conquer rule to identify faulty nodes. It assumes that base station is able to directly transmit messages to any node in the network and rely on other nodes to route measurements to the base station. Also, this approach assumes that each node has a unique identification number. This first step enabled the base station to know the network topology and for this purpose it executes route-discovery protocols. Once the base station knows the node topology it then detects the faulty node by using a simple divide-and-conquer strategy based on adaptive route update messages.

Centralized approach is suitable for certain application. However, it is composed of various limitations. It is not scalable and cannot be used for large networks. Also, due to centralized mechanism all the traffic is directed to and from the central point. This creates communication overhead and quick energy depletions. Moreover, central point is a single point of data traffic concentration and potential failure. Lastly, if a network is portioned, then nodes that are unable to reach the central server are left without any management functionality.

*B. Distributed Approaches*

This is an efficient way of deploying fault management. Each manager controls a sub network and may communicate directly with other managers to perform management functions. Distributed management provides better reliability and energy efficiency and has lower communication cost than centralized management systems [8].

The algorithm proposed for faulty sensor identification in [9] is purely localized. Nodes in the network coordinate with their neighbouring nodes to detect faulty nodes before contacting the central point. In the scheme, the reading of a sensor is compared with its neighbouring' median reading, if the resulting difference is large or large but negative then the sensor is very likely to be faulty. This algorithm can easily be scaled for large network. However, the probability of sensor faults need to be small as this approach works for large networks. Also, if half of the sensor neighbours are faulty and the number of neighbours is even, algorithm cannot detect the fault as expected. But the algorithm developed in [10] tried to overcome the limitations of this approach by identifying good sensor nodes in the network and uses their results to diagnose the faulty nodes. These results are then propagated in the network to diagnose all other sensor nodes. This approach performs well with even number of sensors nodes and do not require sensors physical locations. This approach is not fully dynamic and is required to be pre-configured. Also, each node should have a unique ID and the centre node should know the existence and ID of each node. Another scheme proposed in [11], where sensor nodes police each other in order to detect faults and misbehaviour. Nodes listen-in on the neighbour it is currently routing to and can determine whether the message it sent was forwarded. If the message it sent was not forwarded then it conclude its neighbour as a faulty node and chooses a new neighbour to route to.

The algorithm proposed in [12] is a straightforward and simple mechanism where fault detection is based on the binary output of the sensors. In this approach, each node observes the binary output of its sensor and then compares it with the pre-defined fault model. Fault models can use probability or statistics to detect faulty sensors.

Venkataraman algorithm [10], proposed a failure detection and recovery mechanism due to energy exhaustion. It focused on node notifying its neighbouring nodes before it completely shut down due to energy exhaustion. The paper describes four types of failure recovery mechanisms depending on the type of node in the cluster. The nodes in the cluster are classified into four types, boundary node, pre-boundary node, internal node and the cluster head. Boundary nodes do not require any recovery but pre-boundary node, internal node and the cluster head have to take appropriate actions to connect the cluster. Usually, if node energy becomes below a threshold value, it will send a fail_report_msg to its parent and children. This will initiate the failure recovery procedure so that failing node parent and children remain connected to the cluster.

As we have seen, the distributed approach will be the design trends for fault management in WSNs. Sensor nodes gradually take more management responsibility and decision-making in order to achieve the vision of self-managed WSNs. Node self-detection scheme [13] and neighbour coordination [14] have provided us a good example of management distribution, but their focuses are on a small region (a group of nodes) or individual node. Research work as MANNA [4], WinMS [15] etc proposed management architecture to look after the overall network from a central manager scheme. MANNA [4] is a policy-based approach using external managers to detect faults in the network. MANNA assigns different management roles to various sensor nodes depending on the network characteristics (Homogenous vs. heterogeneous). These distinguish nodes exchange request and response messages with each other for management purpose. To detect node failures, agents execute the failure management service by sensing GET operations for retrieving node states. Without hearing from a node, manager declares it as a faulty node. MANNA has a drawback of providing false debugging diagnosis. There are several reasons a node can be disconnected from the network. It can be disconnected from its cluster and not able to receive any GET message. GET message can be lost during environmental noise. Random distribution and limited transmission range can also cause disconnection. Also, this scheme performs centralized diagnosis and requires an external manager.

WinMS [15] provides a centralized fault management approach. *It uses the central manager with global view of the network to continually analyses network states and executes corrective and preventive management actions according to management policies predefined by human managers.* The central manager detects and localized fault by analyzing anomalies in sensor network models. The central manager analyses the collected topology map and the energy map information to detect faults and link qualities. It has the ability to self configure in case of failure, without prior knowledge of network topology. Also, it analyzes the network state to detect and predict potential failures and perform action accordingly.

## 2.2 Fault diagnosis

In this stage, detected faults are properly identified by the network system and distinguished from the other irrelevant or spurious alarms. Fault diagnosis include fault isolation (where is the fault located), fault identification (what is the type of detected fault), and root cause analysis (what has caused the fault). However, there is still no comprehensive descriptive model to identify or distinguish various faults in WSNs, which supports the network system on accurate fault diagnosis or action-taken in the fault recovery stage [3]. Existing approaches are based on hardware faults and consider hardware components malfunctioning only. Some assume that system software's are already fault tolerant as in [16, 17]. Farinaz [12], described two fault models. The first one corresponds to sensors that produce binary outputs. The second fault model is based on sensors with continuous (analog) or multilevel digital outputs. In [18], the proposed work only consider faulty nodes are due to harsh environment. Thus, there is a need to address a generic fault model that is not based on individual node level, but also consider the network and management aspects.

### 2.3 Failure recovery

In this stage, the sensor network is reconfigured in such a way that failures or faulty nodes do not bring any further impact on the network performance. Most existing approaches isolate faulty (or misbehaving) nodes directly from the network communication layer. For examples, in [11], after the failure of a neighbouring node, a new neighbouring node is selected for routing. WinMS [15], used a proactive fault management maintenance approach i.e. the central manager detect areas with weak network health by comparing the current node or network state with historical network information model (e.g. energy map and topology map). It takes a proactive action by instructing nodes in that area to send data less frequently for node energy consumption. In [19], when a gateway node die, the cluster is dissolved and all its nodes are reallocated to other healthy gateways. This consume more time as all the cluster members are involved in the recovery process. Farinaz [12], suggested a heterogeneous backup scheme for healing the hardware malfunctioning of a sensor node. They believe a single type of hardware resource can backup different types of resources. Although this solution is not directly relevant to fault recovery in respect of the network system level management [3]. In consideration of complexity of fault management design and constrains of a sensor node, we are seeking a localized hierarchical solution to update and reconfigure the management functionality of a sensor node.

In this section, we highlighted different issues and problems existed in already proposed fault management approaches for WSNs. It is clear from the literature survey that different approaches for fault management in WSNs suffer from the following problems:

- Most existing fault management solutions mainly focus on failure detection, and there is still no comprehensive solution available for fault management in WSNs from the management architecture perspective.

- Different mechanisms proposed for fault recovery [12] are not directly relevant to fault recovery in respect of the network system level management i.e. network connectivity and network coverage area etc.

- Failure recovery approaches are mainly application specific, and mainly focus on small region or individual sensor nodes thereby are not fully scalable.
- Some management frameworks require the external human manager to monitor the network management functionalities.

- Another important factor that needs to be considered is vulnerability to message loss. For example, in MANNA [4], if a cluster head does not hear from its cluster member than it announced it as a faulty node. However, a message can be lost due to various reasons. It can be lost during transmission and cause a correct node to be declared as faulty.

We therefore content that there is still a need of a new fault management scheme to address all the problems in existing fault management approaches for wireless sensor networks. We must take into account a wide variety of sensor applications with diverse needs, different sources of faults, and with various network configurations. In addition, it is also important to consider other factors i.e. mobility, scalability and timeliness.

## 3. FAULT MODEL

To facilitate the self managing capability of our proposed fault management scheme, we proposed a new fault knowledge model to support sensor nodes responding to network faults. This knowledge model describes different types of faults for our proposed fault management scheme.

We classified the node fault into two types: permanent, and potential. The permanent fault completely disconnects the sensor node from other nodes, and brings eternal impact on the network performance. For example, hardware faults within a component of a sensor node. A permanent fault once activated remains effective until it is detected and handled. The impact of this failure is usually measured when assessing the network performance. On the other hand, a potential fault usually results from the depletion of node hardware resource, i.e. battery energy. Such fault might cause the node sudden death, and eventually threaten the network life time. When the battery depleted, a node is useless and cannot share in sensing or data dissemination. Potential failure can be detected and treated before it causes the sudden death of a node e.g. sensor node with low residual energy can be send to sleep mode before it completely shuts down and disrupt network operation. Faults can be further classified into: node level fault and network level fault. We proposed a fault model in a tree structure to describe faults monitored in sensor network. As shown in figure 1, "node level" represents the potential and permanent failure of a node while "network level" describes the network faults caused by either potential or permanent failure of one or a set of sensor nodes.

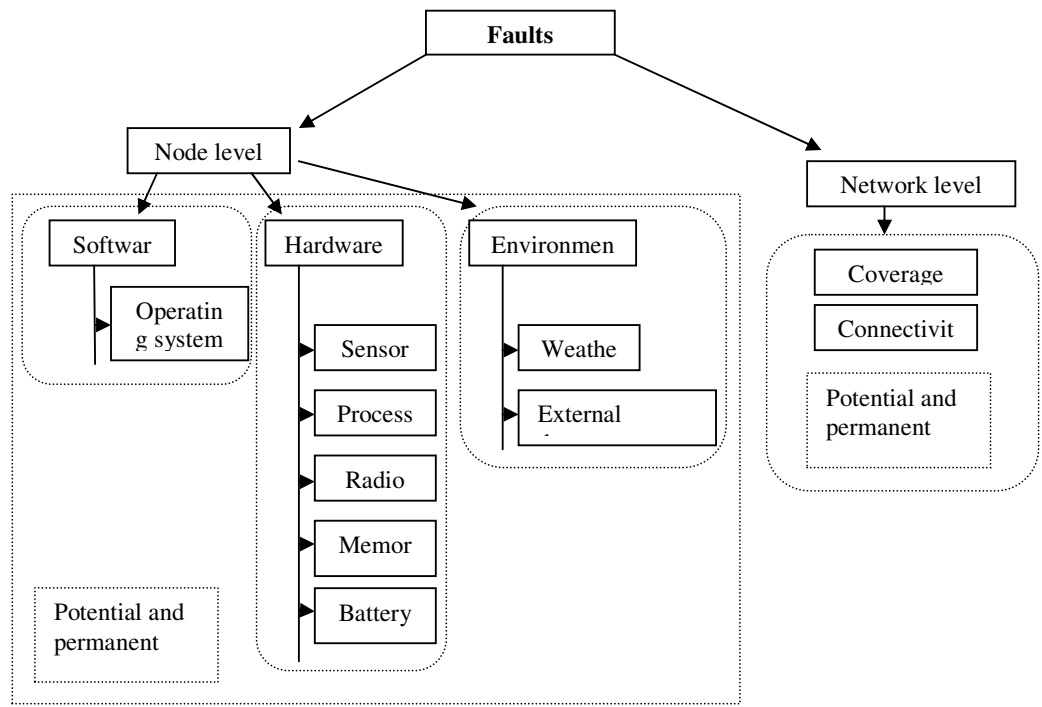

Figure 1. Fault model

Individual node level fault usually results from: application software misbehaviour, hardware failure and external impact of harsh environmental conditions (direct contact with water causing short circuit, node crash by a falling tree etc). In this work, we assume that software components are fault-free or maintained by the sensor application. Fault-tolerance of sensor data have been discussed by various existing research approaches [20]. In this work, we particularly focus on hardware resource depletion as the major cause of sudden death, and its effects at both node and network level. The network level faults are as a result of either the potential or permanent failure, and are usually related to the network connectivity, and sensor coverage rate. In our scheme, the network faults are assessed and analyzed by the management component i.e. group manager, cell manager. It holds the knowledge of its entire region in the network. Based on such information, the fault management system is capable of responding to various network failures with little human administration intervene. For example, when a group manager detect a cell with weak network health, it takes a proactive action by instructing nodes in that cell to send data less frequent for node energy consumption or alternatively, initiate the cell merging procedure.

# 4. A SELF-MANAGING FAULT MANAGEMENT MECHANISM FOR WIRELESS SENSOR NETWORKS

The proposed fault management mechanism can be divided into two phases:

- Fault detection and diagnosis
- Fault recovery

## 4.1 Fault detection and diagnosis

Detection of faulty sensor nodes can be achieved by two mechanisms i.e. self-detection (or passive-detection) and active-detection as shown in figure 2. In self-detection, sensor nodes are required to periodically monitor their residual energy, and identify the potential failure. In our scheme, we consider the battery depletion as a main cause of node sudden death. A node is termed as failing when its energy drops below the threshold value. When a common node is failing due to energy depletion, it sends a message to its cell manager that it is going to sleep mode due to energy below the threshold value. This requires no recovery steps. Self-detection is considered as a local computational process of sensor nodes, and requires less in-network communication to conserve the node energy. In addition, it also reduces the response delay of the management system towards the potential failure of sensor nodes.

To efficiently detect the node sudden death, our fault management system employed an active detection mode. In this approach, the message of updating the node residual battery is applied to track the existence of sensor nodes. In active detection, cell manager asks its cell members on regular basis to send their updates. Such as; the cell manager sends "get" messages to the associated common nodes on regular basis and in return nodes send their updates. This is called in-cell update cycle. The update_msg consists of node ID, energy and location information. As shown in figure 2, exchange of update messages takes place between cell manager and its cell members. If the cell manager does not receive an update from any node then it sends an instant message to the node acquiring about its status. If cell manager does not receive the acknowledgement in a given time, it then declares the node faulty and passes this information to the remaining nodes in the cell. Cell managers only concentrate on its cell members and only inform the group manager for further assistant if the network performance of its small region has been in a critical level.

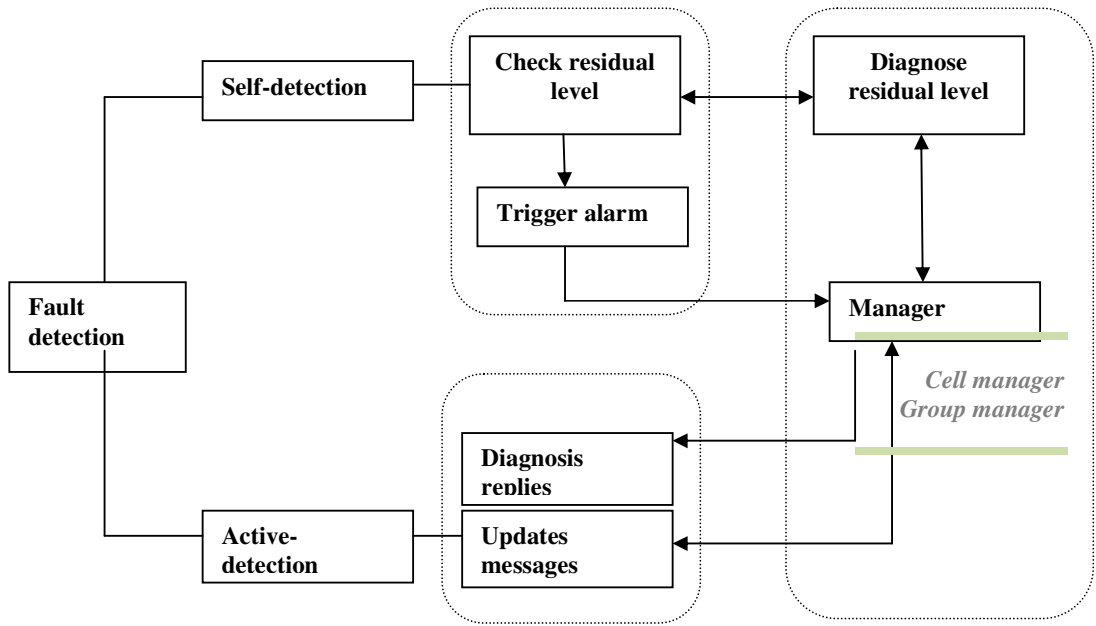

Figure 2. Fault detection and diagnosis process

A cell manager also employs the self-detection approach and regularly monitors its residual energy status. All sensor nodes start with the same residual energy. After going through various transmissions, the node energy decreases. If the node energy becomes less than or equal to 20% of battery life, the node is ranked as low energy node and becomes liable to put to sleep. If the node energy is greater or equal to 50% of the battery life, it is ranked as high and becomes the promising candidate for the cell manager. Thus, if a cell manager residual energy becomes less than or equal to 20% of battery life, it then triggers the alarm and notifies its cell members and the group manager of its low energy status and appoints a new cell manager to replace it.

Every cell manager sends health status information to its group manager. This is called out-cell update cycle and are less frequent than in-cell update cycle. If a group manager does not hear from a particular cell manager during out-cell update cycle, it then sends a quick reminder to the cell manager and enquires about its status. If the group manager does not hear from the same cell manager again during second update cycle, it then declares the cell manager faulty and informs its cell members. This approach is used to detect the sudden death of a cell manager.

Group manager also monitor its health status regularly and respond when its residual energy drops below the threshold value. It notifies its cell members and neighboring group managers of its low energy status and an indication to appoint a new group manager. Sudden death of a group manager can be detected by the base station. If the bases station does not receive any traffic from a particular group manager, it then consults the group manager and asks for its current status. If the base station does not receive any acknowledgement, it then considers the group manager faulty (sudden death) and propagates this information to its cell managers. The base station primarily focuses on the existence of the group managers from their sudden death. Meanwhile, the group managers and cell managers take most parts in passive and active detection in the network.

### 4.2 Fault recovery

After nodes failure detection (as a result of self-detection or active detection), sleeping nodes can be awaked to cover the required cell density or mobile nodes can be moved to fill the coverage hole. A cell manager also appoints a secondary cell manager within its cell to acts as a backup cell manager. Cell manager and secondary cell manager are known to their cell members. If the cell manager energy drops below the threshold value (i.e. less than or equal to 20% of battery life), it then sends a message to its cell members including secondary cell manager. It also informs its group manager of its residual energy status and about the candidate secondary cell manager. This is an indication for secondary cell manager to standup as a new cell manager and the existing cell manager becomes common node and goes to a low computational mode. Common nodes will automatically start treating the secondary cell manager as their new cell manager and the new cell manager upon receiving updates from its cell members; choose a new secondary cell manager. The failure recovery mechanisms are performed locally by each cell. In figure 3, let us assume that cell 1 cell manager is failing due to energy depletion and node 3 is chosen as secondary cell manager. Cell manager will send a message to node 1, 2, 3 and 4 and this will initiate the recovery mechanism by invoking node 3 to stand up as a new cell manager.

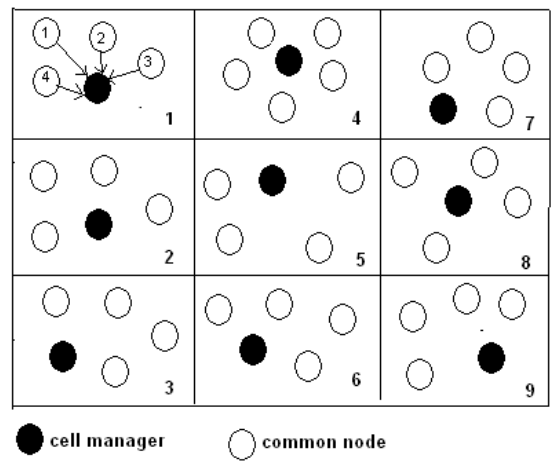

Figure 3. Virtual grid of nodes

In a scenario, where the residual battery energy of a particular cell manager is not sufficient enough to support its management role, and the secondary cell manager also does not have sufficient energy to replace its cell manager. Thus, common nodes exchange energy messages within the cell to appoint a new cell manager with residual energy greater or equal to 50% of battery life. In addition, if there is no candidate node within the cell that has sufficient energy to replace the cell manager. The event cell manager sends a request to its group manager to merge the remaining nodes with the neighbouring cells.

When a group manager detects the sudden death of a cell manager, it then informs the cell members of that faulty cell manager (including the secondary cell manager). This is an indication for the secondary cell manager to start acting as a new cell manager. A group manager also maintains a backup node within the group to replace it when required. If the group manager residual energy drops below the threshold value (i.e. greater or equal to 50% of battery life), it may downgrade itself to a common node or enter into a sleep mode, and notify its backup node to replace it. The information of this change is propagated to neighbouring group managers and cell managers within the group. As a result of group manager sudden death, the backup node will receive a message from the base station to start acting as the new group manager. If the backup node does not have enough energy to replace the group manager, cell managers within a group co-ordinate to appoint a new group manager for themselves based on residual energy.

Each cell maintains its health status in terms of energy. It can be High, Medium or Low. These health statuses are then sent out to their associate group managers periodically during out-cell update cycle. Upon receiving these health statuses, group manager predict and avoid future faults. For example; if a cell has health status high then group manager always recommends that cell for any operation or routing but if the health status is medium then group manager will occasionally recommend it for any operation. Health status Low means that the cell has insufficient energy and should be avoided for any operation. Therefore, a group manager can easily avoid using cells with low health status or alternatively, instruct the low health status cell to join the neighbouring cell. Consider Figure 3, let cell 4 manager is a group manager and it receives health status updates from cell 1, 2 and 3. Cell 2 sends a health status low to its group manager, which alert group manager about the energy status of cell 2.

## 5. MESSAGE BROADCAST ISSUE

The proposed fault management scheme relies on the message exchange among sensor nodes in the network. This might subsequently cause the communication flooding by broadcasting or re-broadcasting messages from different sensor nodes. To address this issue, we employed a message filtering mechanism to further reduce the redundancy of message exchange. The message format contains fields as shown in table 1.

| Group_id | The group id |
|---|---|
| Cell_id | The cell manager id |
| Timestamp | The message sending out time |
| Curr_energy | The current node battery enery |

Table 1. Message attributes

The Group_id field is used to determine whether the received message belongs to the same group of current node. If not, the message will be dropped to avoid unnecessary message re-broadcast. Cell_id field helps a node to decide whether the message belong to its cell. If not, the message will be ignored and not forwarded. A sensor node might receive multiple copies of the same message forwarded by different intermediate nodes. To avoid redundant rebroadcast, we apply the value of 'timestamp' field in the second stage to determine whether the receiving message has been handled previously. If the receiving message is a new one, it will be processed and forwarded to the neighbouring nodes. On the contrary, that message will be dropped to lessen the network traffic and conserve the node energy.

## 6. PERFORMANCE EVALUATION

In this section we evaluate the performance of our proposed algorithm and analyze its cost by measuring node energy expenditure. We used GTSNETS [21] as simulator platform and we used the same radio model as discussed in [22]. In this experiment, we apply fault detection and recovery as main tasks of our fault management approach. Number of sensor is varied from 40 to 80, which are randomly deployed over 120 X 120 square meter area. Each sensor is assumed to have an initial energy of 2000 mJ. Every result shown is an average of 30 experiments. We first compared our work with that of Venkataraman algorithm [23], which is based on failure detection and recovery due to energy exhaustion.

### 6.1 Failure detection

In Venkataraman algorithm, neighboring information is already available to the cluster members through exchange of hello messages. The failure detection procedure starts after the cluster formation. When a node fails, the failing node parents and children take appropriate action to connect the cluster and bridge the gap formed by the failing node. The failing node itself reports its likeliness to fail so that appropriate measures can be taken to rectify the failures. The fail_report-msg is only passed to immediate hop members and then later on passed to the cluster head.

In our proposed algorithm, if node energy drops below a threshold value, it then sends a failure report message directly to its one hop cell manager and goes to a low computational mode. In our proposed algorithm, there are two types of nodes: common node and a cell manager. Only one failure report message is sent out to the cell manager. Thus, avoiding sending any extra message. This reduces the energy consumption and will not disrupt network operation.

### 6.2 Failure recovery

In Venkataraman algorithm, nodes in the cluster are classified into four types: boundary node, pre-boundary node, internal node and the cluster head. Boundary nodes does not require any recovery but pre-boundary node, internal node and the cluster head have to take appropriate actions to connect the cluster. Usually, if node energy becomes below a threshold value, it will send a fail_report_msg to its parent and children. This will initiate the failure recovery procedure so that failing node parent and children remain connected to the cluster. A join_request_mesg is sent by the healthy child of the failing node to its neighbors. All the neighbors within the transmission range respond with a join_reply_mesg/join_reject_mesg messages. The healthy child of the failing node then selects a suitable parent by checking whether the neighbor is not one among the children of the failing node and wether the neighbor is also not a failing node. In our proposed mechanism, common nodes does not require any recovery but goes to low computational mode after informing their cell managers.

In Venkataraman algorithm, cluster head failure causes its children to exchange energy messages. The children who are failing are not considered for the new cluster-head election. The healthy child with the maximum residual energy is selected as the new cluster head and sends a final_CH_mesg to its members. After the new cluster head is selected, the other children of the failing cluster head are attached to the new cluster head and the new cluster head becomes the parent for these children. This cluster head failure recovery procedure consumes more energy as it exchange energy messages to elect the new cluster head. Also, if the child of the failing cluster head node is failing as well, then it also requires appropriate steps to get connected to the cluster. These can disrupt network operation and is time consuming.

In our proposed algorithm, we employ a back up secondary manager which will replace the cell manager in case of failure. Every time a cell manager is failing it sends a message to all its members including the backup secondary cell manager. Upon receiving this message from its cell manager, secondary manager automatically starts acting as a new cell manager and no further messages are required to send to other cell members to inform them about the new cell manager as they are already aware of secondary cell manager.

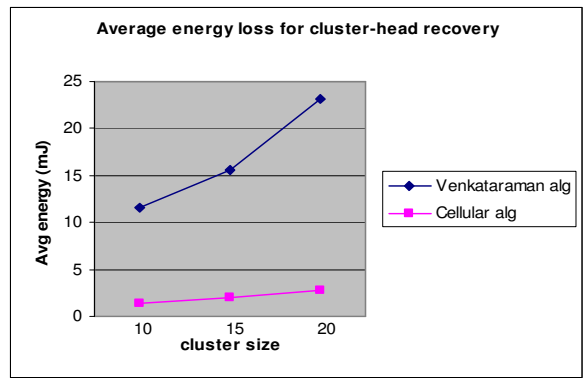

Figure 4. Average energy loss for cluster head recovery

It can be observed from figure 4 that our proposed algorithm consumes less energy for cluster head failure recovery when compared to Venkataraman algorithm. In Venkataraman algorithm, message exchange for the election of new cluster manager is both time and energy consuming. In our proposed algorithm, cell manager sends one message only to its member to recover from a failure.

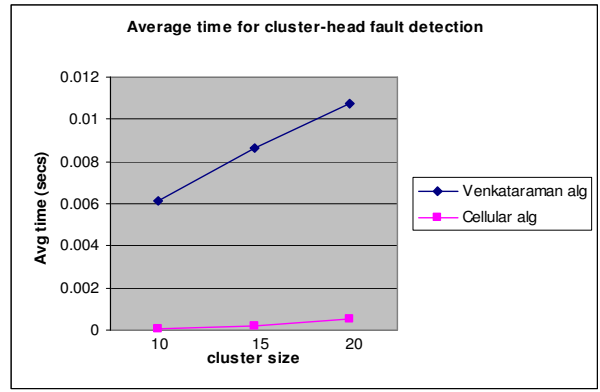

Figure 5. Average time for cluster head recovery

Figure 5 depicts the average time required for the cluster head recovery. It can be observed that our proposed algorithm perform a quicker recovery as compared to Venkataraman algorithm.
We also compared our scheme with two other algorithms: autonomic self-organizing architecture [24] and load- balanced clustering [22], in terms of energy consumption for cluster head recovery. It can be observed from figure (6) that our proposed algorithm consumes less energy in re-clustering when compared to the other two.
In autonomic self-organizing algorithm, when a high level node (header) failed to operate or need to step down due to low residual energy. All sensor nodes from the failed header need to join other available header nodes using the same mechanism. This again is not an energy efficient way to re-organize the cluster and also time consuming as compared to our cellular approach. In load-balanced clustering, when a gateway fails, the cluster dissolved and all its nodes are re-allocated to other healthy gateways. This consumes more time and energy as all cluster members are involved in the re-clustering process. In our proposed algorithm, only few nodes are involved in re-clustering.

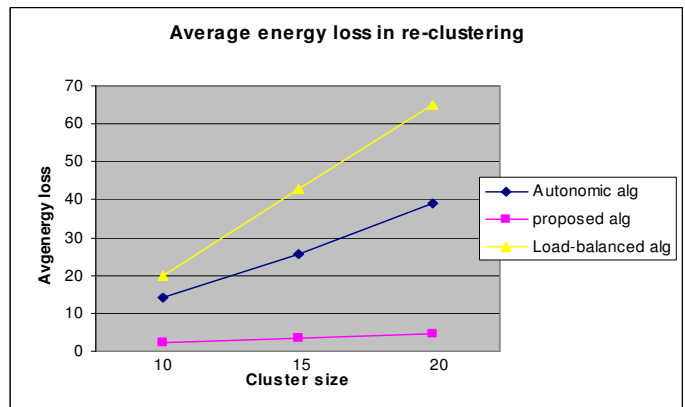

Figure 6. Average energy loss in re-clustering

## 7. SUMMARY

Wireless sensor network are composed of many wireless sensing devices called sensor nodes. These nodes are small in size, limited in resources and randomly deployed in harsh environment. Therefore, it is not uncommon for sensor networks to have malfunction behaviour, node, link or network failure. In this paper, we proposed a fault management mechanism for wireless sensor network to diagnose faults and perform appropriate measures to recover sensor network from failures. The proposed fault management mechanism is energy-efficient and responsive to network topology. We proposed a fault model that describes different types of faults at different levels of the hierarchy. Depending on the role assignment, sensor nodes execute the appropriate functions to complete their fault management tasks. Most of existing solution used some type of central entity to perform fault management tasks but in our proposed solution, the aim is to perform fault detection locally and in distributed fashion. The result obtained from the simulation clearly shows that our proposed algorithm performs failure detection and recovery much faster than other existing schemes, and consumed significantly lower energy.

## 8. REFERENCES

[1] M. Z. Khan, M. Merabti, and B. Askwith, "Design Considerations for Fault Management in Wireless Sensor Networks," in *PGNet 2009* Liverpool, 2009.
[2] L. Paradis and Q. Han, "A Survey of Fault Management in Wireless Sensor Networks," *Journal of Network and Systems Management,* vol. 15, pp. 171-190, 2007.
[3] M. Yu, H. Mokhtar, and M. Merabti, "A survey on Fault Management in wireless sensor network," in *Proceedings of the 8th Annual PostGraduate Symposium on The Convergence of Telecommunications, Networking and Broadcasting* Liverpool, UK, 2007.
[4] L. B. Ruiz, I. G.Siqueira, L. B. Oliveira, H. C. Wong, J. M. S. Nogueira, and A. A. F. Loureiro, "Fault management in event-driven wireless sensor networks," in *MSWiM'04* Italy, 2004.
[5] M. Asim, H. Mokhtar, and M. Merabti, "A cellular approach to fault detection and recovery in wireless sensor networks," in *The Third International Conference on Sensor Technologies and Applications, SENSORCOMM 2009* Greece, 2009.
[6] N. Ramanathan, K. Chang, E. Kohler, and D. Estrin, "Sympathy for the Sensor Network Debugger," in *Proceedings of 3rd ACM Conference on Embedded Networked Sensor Systems (SenSys '05),* San Diego, California, 2005, pp. 255-267.
[7] J. Staddon, D. Balfanz, and G. Durfee, "Efficient Tracing of Failed Nodes in Sensor Networks," in *First ACM International Workshop on Wireless Sensor Networks and Applications* USA, 2002.
[8] W. L. Lee, A. Datta, and R. Cardell-Oliver, "Network Management in Wireless Sensor Networks," in *Handbook of Mobile Ad Hoc and Pervasive Communications*: American Scientific Publishers, 2006.
[9] M. Ding, D. Chen, K. Xing, and X. Cheng, "Localized fault-tolerant event boundary detection in sensor networks," in *Proceedings of the 24th Annual Joint Conference of the IEEE Computer and Communications Societies (INFOCOM '05)*. vol. 2 USA, 2005, pp. 902-913.
[10] J. Chen, S. Kher, and A. K. Somani, "Distributed Fault Detection of Wireless Sensor Networks," in *Proceedings of DIWANS 06*, 2006.
[11] S. Marti, T. J. Giuli, K.Lai, and M. Baker, "Mitigating routing misbehaviour in mobile ad hoc networks," in *ACM Mobicom*, 2000, pp. 255-265.
[12] F. Koushanfar, M. Potkonjak, and A. SangiovanniVincentelli, "Fault tolerance techniques in wireless ad-hoc sensor networks," UC Berkeley technical reports 2002.
[13] A. R. S Harte, K M Razeeb, "Fault Tolerance In Sensor Networks using Self-Diagnosing Sensor Nodes."
[14] M. L. Chihfan Hsin, "Self-monitoring of Wireless Sensor Networks," *Computer Communications,* vol. 29, pp. 462-478, 2005.
[15] W. L. Lee, A. Datta, and R. Cardell-Oliver, "WinMS: Wireless Sensor Network-Management System, An Adaptive Policy-Based Management for Wireless Sensor Networks," School of

Computer Science and Software Engineering, The University of Western Australia, Technical Report UWA-CSSE-06-01, 2006.

[16]  J. Chen, S. Kher, and A. Somani, "Distributed Fault Detection of Wireless Sensor Networks," in *DIWANS'06* USA, 2006.

[17]  F. Koushanfar, M. Potkonjak, and A. Sangiovanni-Vincentelli, "Fault Tolerance in Wireless Ad-hoc Sensor Networks," in *Proceedings of IEEE Sensors*, 2002.

[18]  T. Clouqueur, K.Saluja, and P. Ramanathan, "Fault Tolerance in Collaborative Sensor Networks for Target Detection," in *IEEE Transactions on Computers*, 2004, pp. 320-333.

[19]  G. Gupta and M. Younis, "Fault-Tolerant Clustering of Wireless Sensor Networks," in *Proceedings of the IEEE WCNC 2003* New Orleans, Louisiana, 2003.

[20]  K. F. Ssu, C. H. Chou, H. C. Jiau, and W. T. Hu, "Detection and Diagnosis of data inconsistency failures in wireless sensor networks," in *Computer Networks*, 2006, pp. 1247-1260.

[21]  G. Riley, "The Georgia Tech Network Simulator," in *ACM SIGCOMM Workshop on Models, Methods and Tools for Reproducible Network Research* Karlsruhe, Germany, 2003.

[22]  G. Gupta and M. Younis, "Load-Balanced Clustering in Wireless Sensor Networks," in *Proceedings of International Conference on Communication (ICC 2003)* Anchorage, AK, 2003.

[23]  G. Venkataraman, S. Emmanuel, and S.Thambipillai, "Energy-efficient cluster-based scheme for failure management in sensor networks," in *IET Communications*. vol. 2, 2008, pp. 528-537.

[24]  J. L. Chen, H. F. Lu, and C. A. Lee, "Autonomic self-organization architecture for wireless sensor communications," *International Journal of Network Management,* vol. 17, pp. 197-208, 2007.

## Authors

1.  **Muhammad Asim**

    P.h.D Student, Liverpool John Moores University, UK

2.  **Dr. Hala Mokhtar**

    Senior lecturer, Department of Networked Systems and Security, Liverpool John Moores University, UK

3.  **Professor Madjid Merabti**

    Professor Madjid Merabti is Director at the School of Computing & Mathematical Sciences, Liverpool John Moores University, UK.